# Liquid-metal/NdFeB/Sillicone Composite Elastomer with Reprogrammable Magnetization and Modulus

Ran Zhao[a,b,*], Guopeng Zhou[c], Hanchen Yao[a], Houde Dai[a,*]

[a] Quanzhou Institute of Equipment Manufacturing of Haixi Institutes, Chinese Academy of Sciences, Quanzhou 362000, P.R. China;

[b] Zhongyuan-Petersburg Aviation College, Zhongyuan University of Technology, Zhenzhou 45000, P.R. China;

[c] Zhongbei University, Taiyuan 030000, P.R. China;

[*]E-mail: Ran Zhao (*zhaoran@zut.edu.cn*); Houde Dai(*dhd@fjirsm.ac.cn*)

**Abstract:** Magnetic programming soft machines has great development prospects in the fields of minimally invasive medicine, wearable device and soft robot. However, unrepeatable magnetization and low modulus limits their applications. So far, there are few techniques that can make magnetic soft robots have adjustable functions, mechanical and electrical properties. This paper presented a magnetic functional elastomer based on Liquid-metal/NdFeB/Sillicone composite materials. This magnetoelastomer uses the phase transition characteristics of liquid-metal and magnetic guided rotation of micro ferromagnetic particles to realize reprogrammable magnetization and modulus. The elastomer's fabrication method was given and the basic programmable properties were tested. Some robot prototypes have also been manufactured to show how to respond to different mission requirements by programming magnetization and modulus. Our research provides a new path for the design and large-scale manufacturing of magnetic soft robot.

## 1. Introduction

Magneto actuated soft robots have the advantages of fast deformations, unlimited endurance, no obstruction restrictions, and magnetic tracking [1-3], comparing with light-, heat-, or chemo-actuated soft robots [4-6]. But most of them are single-function machines[1]. At present, the technical limitations of magnetic soft robot are mainly in two aspects: Firstly, their function cannot change. Secondly, their stiffness are too low to implement the intervention or load carrying tasks[7-9].

So far, there are not many effective technologies to realize the repeatable magnetic programming of soft robots and change their stiffness. Sitti *et al.* developed a heat-assisted reprogrammable magnetization method—to change the magnetization profiles by heating/cooling the $CrO_2$ particles to above/below the Curie temperature (118°C) [10]. Lin and coworkers adopted shape memory polymers (SMPs) as a flexible matrix and re-patterned magnetic particles based on the thermally induced glass-transition principle [11]. The stiffness control methods of soft robots mainly rely on the change of the Young's modilli caused by the glass-transition characteristics of the SMPs. Zhao et al. utilized $Fe_3O_4$/SMP composites to perform stiffness adjustments of soft structures through heating induced by a high-frequency magnetic field [12] (the conversion temperature is 65°C). Leng *et al.* developed a flexible 3D structure based on Fe3O4/PLA composites, its stiffness and deformation are controllable with a 160°C glass transition temperature [13]. However, the change range of the Young's modulus caused by the glass transition between the stationary and reversible phase of the SMPs was limited (2-100 times) [14-16]. The control temperature of these techniques were far beyond the tolerable temperature for organisms. It is still a great challenge to implement these two technologies simultaneously on untethered soft robots.

Therefore, we developed a functional elastomer based on the liquid-metal/NdFeB/Sillicone composites. The proposed elastomer has programmable magnetization and modulus. The details are given as follow.

## 2. Principle

The designed elastomer has the characteristic of programmable magnetization and modulus, which is based on the phase transition property of Liquid metal and the reconfigurable polarity pole of ferromagnetic liquid metal plasticine (LM-MFs). Fig.1 shows the working principle of the proposed functional elastomer.

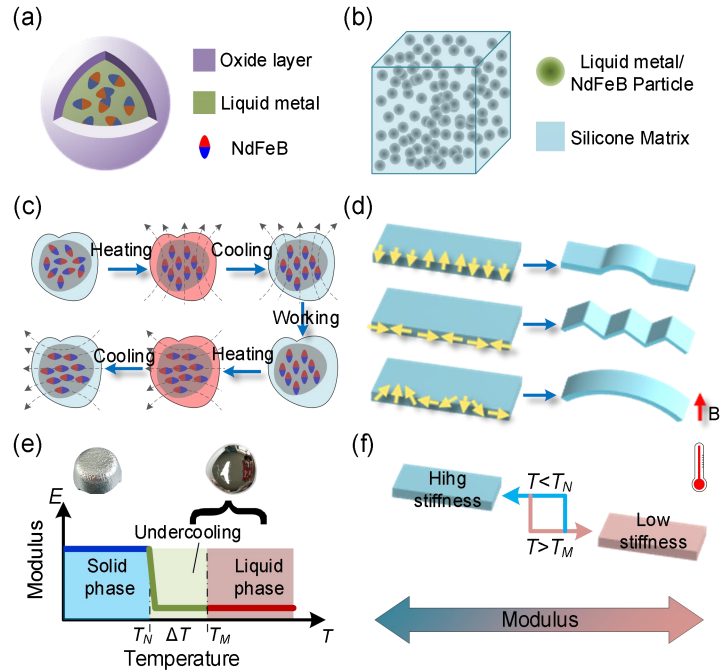

**Figure. 1** Principle of the proposed programmable elastomer: (a) Micro structure of LM-FMs particle, (b) Structure the composite elastomer, (c) Reprogramming magnetization mechanism of the liquid-metal/NdFeB/silicone micro cell, (d) Magnetization profiles and shape morphing, (e) Phase transition characteristics of liquid metal, and (f) Programmable modulus/stiffness change of the elastomer based on liquid-metal phase switching.

Fig.1a exhibits the micro structure of a LM-FMs composites particle. Its exterior is a metal oxide layer ( which is $Ga_2O_3$ for gallium based Liquid metal), and the core is the functional part－NdFeB particles wrapped by Liquid metal. The elastomer is composed of silicone matrix and Liquid metal/NdFeB magnetic particles (shown in Fig.1b).

Fig.1c describes the magnetization process of a micro cell: (a) first, convert the liquid metal from solid to liquid state by heating; (b)then aligned the NdFeB hard magnetic particles by using an external magnetic field; (c) at last, stop heating and the liquid metal can lock NdFeB particles. And the process is repeatable.It should be pointed out that due to the high rheological properties of liquid metal magnetic plasticine, when the liquid metal is in the liquid phase, it can still lock the NdFeB particles until it is reprogrammed by a strong external magnetic field. This characteristic has been proved by Ref.[x] and our previous work. This means that when the driving magnetic field is less than the value of programmed magnetic field, the programmed magnetization direction will not change regardless of which phase the liquid metal is. Fig.1d shows that different magnetic response actions can be realized by programming different magnetization profiles on the film.

The phase transition theory of liquid metal is given in Fig.1e. There are three states of liquid metal, including solid phase state, liquid phase state and undercooling state. Between the melting point ($T_M$) and the solidification point ($T_N$), the liquid metal can usually remain liquid in a certain temperature range, which is called undercooling and defined as,

$$\Delta T = T_M - T_N \tag{1}$$

($\Delta T$). This phenomenon requires us to realize the phase switching of liquid metal at an ambient temperature below the solidification point.

$$\Delta T \leq 0.18 T_M \tag{2}$$

Fig.1f demonstrates that how to programming stiffness of the elastomer by adjusting the temperature. The equivalent Young's modulus of the magnetic elastomer can be solved by using the Voigt limits [17], which is expressed as,

$$E_e = \alpha_1 E_{LM-FMs} + \alpha_2 E_{silicone} \tag{3}$$

where $E_{LM-MFs}$ and $E_{silicone}$ are the Young's modulus of LM-FMs and silicone, respectively; $\alpha_1$ and $\alpha_2$ are the volume fractions of LM-FMs inclusion and silicone matrix, and there is $\alpha_1 + \alpha_2 = 1$.

## 3. Materials and Method

### 3.1 Preparation of liquid metal/NdFeB composites particle

The $Nd_2Fe_{14}B$ powder used here has average size of 6 μm (Xinnuode Transmition Devices Technology co Ltd, China). Two kind of liquid metal with phase transition temperature of 25°C (Ga-In-Sn alloy) and 30°C ( pure gallium) were used as the matrix. The liquid metal was heated to 50 degrees and then it is mixed with Nd-Fe-B particles in the volume ratio of 6:4. The mixture was put into a mechanical agitator and stirred in air environment for 10 minutes. The liquid metal was solidified under a low temperature and then the bulk was soaked in ethanol solution and ball milled to obtain micron particles. As shown in Fig.1a, due to exposure into the air, the particles will form a thin $Ga_2O_3$ shell, which can protect the internal liquid metal from further oxidation.Besides, this oxide layer can decrease the surface surface tension of liquid metal/NdFeB composites, and make it easier to mix with silicone.We also propose a template-assisted method to prepare sub-millimeter size particles (100~800 μm). With the help of template, particles with special shapes, such as needles, can be made. These special shaped particles are helpful to regulate the properties of magnetic elastomer.

**3.2 Fabrication of magnetic elastomer and soft robots**

Figs.2a and 2b shows the preparation process of the magnetic elastomer. The liquid metal/NdFeB composite particles and silicone (Echoflex-50, Smooth on, America) were mixed in a ratio of 6:4. Then the mixture was put into a PTFE mold. The mold and mixture were placed in a vacuum heating box, vacuumized, defoaming, and heated at 30°C for 30 minutes, then exposed to air for 2 hours for curing. As shown in Fig.2c the elastomer was put into an electromagnetic solenoid, and magnetized by a 1.5 T magnetic field.

The functional elastomer can be used to fabricate magnetic soft robot with various shapes. Mechanical cutting (Waterjet Cutting, Die cutting) or laser cutting methods can be used to obtain the designed contour (shown in Fig.2d). The dimensions of these soft robots range from microns to centimeters.

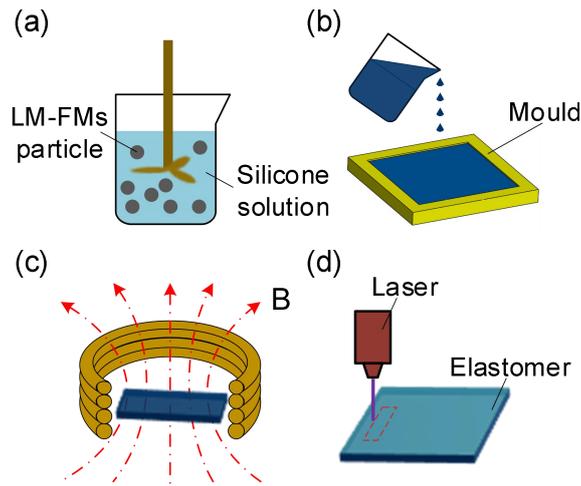

Fig.2 The fabrication process of the elastomer and robot

## 4. Results and discussions

### 4.1 Programmable magnetization

Fig.3 shows the magnetic field induced shape-morphing of the robots. Three kind of robots were designed, and programmed at least two magnetization profiles. In fig.3a, the robot shows "L", "Λ" and twisting shape when programmed magnetization profile I, II and III. As shown in fig.3b, the robot shows "S", "Long L" and curling shape when programmed magnetization profile IV, V and VI. In fig.3c, the robot performs standing and grasping actions when programmed magnetization profile VII, and VII.

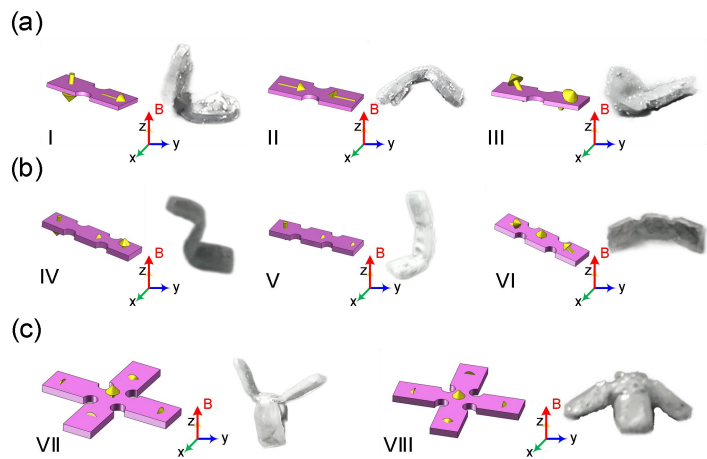

Fig.3 Magnetic responses of different programmable profiles

### 4.2 Programmable modulus

Fig. 4 shows the switching times of the robot between soft-body and rigid-body forms when the ambient temperature was 22.3°C. In the heating process, the robot was heated to 50, 40, and 35°C. The switching time increased linearly with the heating rate. In the cooling process, the robot was cooled to 0 and 22.3°C by electronic cooling and natural cooling methods, respectively. According to Equation (2), to ensure the solidification of the LM, the refrigeration temperature difference should be set to be larger than 5.4°C.

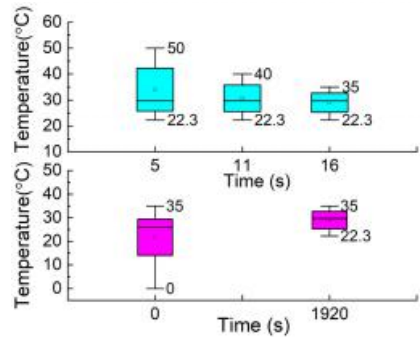

Fig.4 The transformation time of the heating and cooling processes

The calculated and experimental value of elastic modulus of the elastomer are given in Fig.5, When converting to rigid body form, the modulus increases by 2000 times.

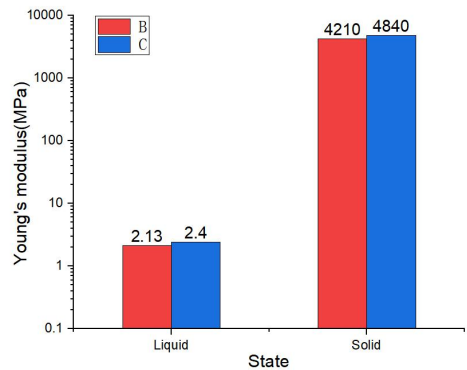

Fig.5 The Young's modulus of the elastomer

**4.3 Discussions**.

The presented reprogrammable magnetization technique decouples the processes of robot manufacturing and magnetic programming and introduces the unique property of a reconfigurable function to the magnetic soft robot. Compared with the methods presented in Refs.[17] and [18], the new programming method requires low energy.

According to a calculation, when the robot switches from soft-body form to rigid-body form, its stiffness increases 2000 times. This variation range is much larger than those of the variable stiffness robots previously reported [19].

## 5. Conclusion

This paper presents a functional elastomer for magnetic soft robots. A new magnetic response materials of liquid metal/NdFeB/Silicone composite was developed to fabricate the elastomer. A batch manufacturing method of programmable magnetic soft robot was also proposed. Some robots are designed to testify the function of the elastomer.

This research provides a new idea for the design of magnetic soft robots. Future work will focus on producing intelligent soft robots with 3D structures.

**Declaration of Competing Interest**

The author declares that they have no known competing financial interests or personal relationships that could have appeared to influence the work reported in this paper.